\documentclass[twoside,11pt]{article}
\usepackage[accepted]{melba} 
\usepackage[utf8]{inputenc}
\usepackage[T1]{fontenc}
\usepackage{amsmath}
\usepackage{amssymb}
\usepackage{relsize}
\usepackage{algorithm} 
\usepackage{algpseudocode} 
\usepackage{subfig}
\usepackage{graphicx}
\usepackage{subfig}
\usepackage{cancel}

\usepackage{changepage}

\usepackage[dvipsnames]{xcolor}

\DeclareMathOperator{\EX}{\mathbb{E}}
\usepackage{url}
\usepackage{subfig}
\usepackage{graphicx}
\usepackage{subfig}
\usepackage[flushleft]{threeparttable}
\usepackage{changepage}
\usepackage{cancel}

\melbaheading{012}{https://www.melba-journal.org/article/27648}{2021}{1-25}{11/2020}{09/2021}{Dalca and Sabuncu}{}{}

\ShortHeadings{Unsupervised Detection of Hypoplastic Left Heart Syndrome}{}
\firstpageno{1}

\title{Learning normal appearance for fetal anomaly screening: Application to the unsupervised detection \\ of Hypoplastic Left Heart Syndrome}

\author{\name Elisa Chotzoglou \email e.chotzoglou16@imperial.ac.uk \\ 
\addr  Imperial College London, London, UK 
\AND
\name Thomas Day \email thomas.day@kcl.ac.uk \\
\addr King's College London, UK
\AND
\name Jeremy Tan \email j.tan17@imperial.ac.uk \\
\addr  Imperial College London, London, UK 
\AND
\name Jacqueline  Matthew \email jacqueline.matthew@kcl.ac.uk\\
\addr King's College London, UK
\AND
\name  David  Lloyd \email david.lloyd@kcl.ac.uk\\
\addr King's College London, UK
\AND
\name  Reza  Razavi \email reza.razavi@kcl.ac.uk\\
\addr King's College London, UK
\AND
\name  John  Simpson \email John.Simpson@gstt.nhs.uk\\
\addr King's College London, UK
\AND
\name  Bernhard  Kainz \email b.kainz@imperial.ac.uk\\
\addr  Imperial College London, London, UK 
}

\begin{document}
\maketitle


\begin{abstract}
Congenital heart disease is the most common group of congenital malformations, affecting $6-11$ per $1000$ newborns. In this work, an automated framework for detection of cardiac anomalies during ultrasound screening is proposed and evaluated on the example of Hypoplastic Left Heart Syndrome (HLHS), a sub-category of congenital heart disease. We propose an unsupervised approach that learns healthy anatomy exclusively from clinically confirmed normal control patients. We evaluate a number of known anomaly detection frameworks together with a model architecture based on the $\alpha$-GAN network and find evidence that the proposed model performs significantly better than the state-of-the-art in image-based anomaly detection, yielding  average $0.81$ AUC \emph{and} a better robustness towards initialisation compared to previous works.  
\end{abstract}

\begin{keywords}
fetal screening, detection, unsupervised learning
\end{keywords}

\section{Introduction}
A contemporary key element in building automated pathology detection systems with machine learning in medical imaging is the availability and accessibility of a sufficient amount of data in order to train supervised discriminator models for accurate results. This is a problem in medical imaging applications, where data accessibility is scarce because of regulatory constraints and economic considerations. To build truly useful diagnostic systems, supervised machine learning methods would require a large amount of data and manual labelling effort for every possible disease to minimise false predictions. This is unrealistic because there are thousands of diseases, some represented only by a few patients ever recorded. 
Thus, learning representations from healthy anatomy and using anomaly detection to flag unusual image features for further investigation defines a more reasonable paradigm for medicine, especially in high-throughput settings like population screening, e.g. fetal ultrasound imaging. However, anomaly detection suffers from the great variability of healthy anatomical structures from one individual to another within patient populations as well as from the many, often subtle, variants and variations of pathologies. Many medical imaging datasets, e.g. volunteer studies like UK Biobank \citep{petersen2013imaging}, consist of images from predominantly healthy subjects with a small proportion of them belonging to abnormal cases. Thus, an anomaly detection approach or 'normative' learning paradigm is also reasonable from a practical point of view for applications like quality control within massive data lakes. 

In this work, we formulate the detection of congenital heart disease as an anomaly detection task for fetal screening with ultrasound imaging. We utilise normal control data to learn the normative feature distribution which characterises healthy hearts and distinguishes them from fetuses with hypoplastic left heart syndrome (HLHS). We chose this test pathology because of our access to a well labelled image database from this domain. Theoretically, our method could be evaluated on any congenital heart disease that is visible in the four-chamber view of the heart~\citep{screening2015}. 

\textbf{Contribution:} 
To the best of our knowledge we propose the first unsupervised working anomaly detection approach for fetal ultrasound screening using only normal samples during training. Previous approaches rely on supervised~\citep{similarwork:2020} discrimination of known diseases, which makes them prone to errors when confronted with unseen classes. Our method extends the $\alpha$-GAN architecture with attention mechanisms and we propose an anomaly score which is based on reconstruction and localisation capabilities of the model. We evaluate our method on a selected congenital heart disease, which can be overlooked during clinical screening examinations in between 30-40\% of scans~\citep{chew2007population}, and compare to other state-of-the-art methods in image-based anomaly detection. We show evidence that the proposed method outperforms state-of-the-art models and achieves promising results for unsupervised detection of pathologies in fetal ultrasound screening.  

\section{Background and Related Work}

\subsection{Pathological Diseases in Fetal Heart}
Congenital heart disease (CHD) is the most common group of congenital malformations \citep{chd1:2010}\citep{chd2:2018}\citep{chd3:2016}. CHD is a defect in the structure of the heart or great vessels that is present at birth. Approximately $6-11$ per $1000$ newborns are affected. $20-30\%$ of these heart defects require  surgery within the first year of life~\citep{chd2:2018}. In order to detect the disease, the most common approach is the standard anomaly ultrasound scan at approximately $20$ weeks of gestation (e.g. 18+0 to 20+6 weeks in the UK). In contemporary screening pathways, \emph{i.e.}, 2D ultrasound at GA 12 and 24, the prenatal detection rate  of CHD is in a range of $39-59\%$. \citep{bariers:2012} \citep{chd3:2016}
In \citep{chd2:2018}, algorithmic support has been used to find diagnostically informative fetal cardiac views. With this aid, clinical experts have been shown to discriminate healthy controls from CHD cases with $98\%$ sensitivity and $90\%$ specificity in 4D ultrasound. However, 4D ultrasound is not commonly used during fetal screening and in the proposed teleradiology setup still all images have to be manually assessed by highly experienced experts to achieve such a high performance. 

\begin{figure}[ht!]
\centering
\includegraphics[height=4cm]{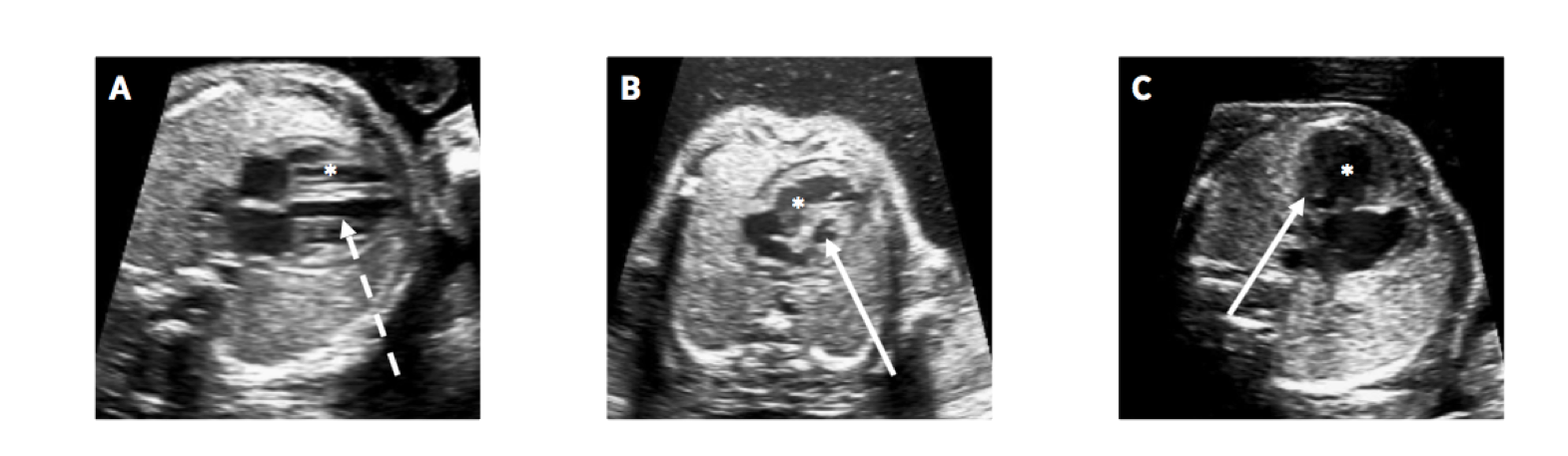}
\caption{Examples of four-chamber views of the fetal heart. A shows a normal fetal heart, with the normal sized LV (left ventricle) marked (dashed white arrow). B and C show two examples of fetal HLHS (hypoplastic left heart syndrome), with the hypoplastic LV marked (solid white arrow). Example B represents the mitral stenosis / aortic atresia subtype, with a severely hypoplastic, globular LV. Example C represents the mitral atresia / aortic atresia subtype, with a slit-like LV that is difficult to identify.  * marks the right ventricle in each case.}
\label{fig:normal_abnormal}
\end{figure}

In this work we focus on a subtype of CHD, Hypoplastic Left Heart Syndrome (HLHS).
Examples of HLHS in comparison with healthy fetal hearts are presented in
Figure~\ref{fig:normal_abnormal}. 
HLHS is rare, but is one of the most prominent pathologies in our cohort.  In HLHS the four chamber view is usually grossly abnormal, allowing the identification of CHD (although not necessarily a detailed diagnosis) from a single image plane. A condition that is identifiable on a single view plane provides a clear case study for our proposed method. If HLHS is identified during pregnancy, provisions for the appropriate timing and location of delivery can be made, allowing immediate treatment of the affected infant to be instigated after birth. Postnatal palliative surgery is possible for HLHS, and the antenatal diagnosis of CHD in general has been shown to result in a reduced mortality compared to those infants diagnosed with CHD only after birth \citep{holland:2015}. However, the detection of this pathology during routine screening still remains challenging. Screening scans are performed by front-line-of-care sonographers with varying degrees of experience and the examination is influenced by factors such as fetal motion and the small size of the fetal heart.

\subsection{One-class anomaly detection methods in Medical Imaging}
\label{sotamethods}
One-class classification is a case of multi-class classification where the data is from a single class. The main goal is to learn either a representation or a classifier (or a combination of both) in order to distinguish and recognise out-of-distribution samples during inference. Discriminative as well as generative methods have been proposed utilizing deep learning, for example one class CNN~\citep{uncertainty110:2019} and Deep SVDD~\citep{deepsvdd:2018}. Usually these methods utilise loss functions, similar to those of OC-SVM~\citep{uncertainty108:2001}  and SVDD~\citep{uncertainty109:2004} or use regularisation techniques to make conventional neural networks compatible to one-class classification models~\citep{uncertainty107:2021}.
Generative models are mostly based on autoencoders or Generative Adversarial Networks. In this work we mainly focus on the application of generative adversarial networks for anomaly detection in medical imaging.

Generative adversarial networks for anomaly detection were first proposed by~\citep{anogan:2017}. In \citep{anogan:2017}, a deep convolutional  generative adversarial network, inspired by DCGAN as proposed by \citep{dcgan:2016}, is used as \emph{AnoGAN}. During the training phase, only healthy samples are used. This approach consists of two models. A generator, which generates an image from random noise and a discriminator, which classifies real or fake samples as common in GANs. More specifically, the generator learns the mapping from the uniformly distributed input noise sampled from the latent space to the 2D image space of healthy data. The output of the discriminator is a single value, which is interpreted as the probability of an image to be real or generated by the generator network. In their work, a residual loss is introduced, which is defined as the $l1$ norm between the real images and the generated image. This enforces the visual similarity between the initial image and the generated one. Furthermore, in order to cope with GAN instability, instead of optimizing the parameters of the generator via maximizing the discriminator’s output on generated examples, the generator is forced to generate data whose intermediate feature representation of the discriminator ($D_{H}$) is similar to those of real images. This is defined as the $l1$ norm between intermediate feature representations of the discriminator given as input the real image and the generate image respectively. In AnoGAN, an anomaly score is defined as the loss function at the last iteration, \emph{i.e.}, the residual error plus the discrimination error. AnoGAN has been tested on a high-resolution SD-OCT dataset. For evaluation purposes, the authors report receiver operating characteristic (ROC) curves of the corresponding anomaly detection performance on image level. Based on their results, using the residual loss alone already yields good results for anomaly detection. The combination with the discriminator loss improves  the overall performance slightly. During testing, an iterative search in the latent space is used in order to find the closest latent vector that reconstructs the  real test image better. This is a time consuming procedure and this optimisation process can get stuck in local minima. 

Similar to AnoGAN, a faster approach, f-AnoGAN has been proposed in \citep{fanogan:2019}. In this work, the authors train a GAN on normal images, however instead of the DCGAN model a Wasserstein GAN (WGAN) \citep{wgan:2017}\citep{improvedwgan:2017}  has been used. Initially, a WGAN is trained in order to learn a non-linear mapping from latent space to the image space domain. Generator and discriminator are optimised simultaneously. Samples that follow the data distribution are generated through the generator, given input noise sampled from the latent space. Then an encoder (convolutional autoencoder) is training to learn a map from image space to latent space. For the training of the encoder, different approaches are followed, i.e  training an encoder with generated images (z-to-z approach-\textit{ziz}), training an encoder with real images (an image-to-image mapping approach -\textit{izi}) and training a discriminator guided izi encoder (\textit{$izi_{f}$}). As anomaly score, image reconstruction residual plus the residual of the discriminator's feature representation ($D_{H}$) is used. The method is evaluated on optical coherence tomography imaging data of the retina. Both \citep{anogan:2017} as well as \citep{fanogan:2019} use image patches for training and are modular methods which are not trained in an end-to-end fashion. 

Another GAN-based method applied to OCT data has been proposed by~\citep{isbi:2020}, in which authors propose a Sparsity-constrained Generative Adversarial Network (Sparse-GAN), a network based on an Image-to-Image GAN~\citep{imagetoimage:2017}. Sparse-GAN consists of a generator, following the same approach as in \citep{imagetoimage:2017}, and a discriminator. Features in the latent space are constrained using a Sparsity Regularizer Net. The model is optimized with a reconstruction loss combined with an adversarial loss. The anomaly score is computed in the latent space and not in image space. 
Furthermore, an Anomaly Activation Map (AAM) is proposed to visualise lesions.

Subsequently, AnoVAEGAN \citep{baur2018deep:2018} has been proposed, in which the authors discuss a spatial variational autoencoder and a discriminator. It is applied to high resolution MRI images for unsupervised lesion segmentation. AnoVAEGAN uses a variational autoencoder and tries to model the normal data distribution that will lead the model to fully reconstruct the healthy data while it is expected to fail reconstructing abnormal samples. The discriminator classifies the inputs as real or reconstructed data. As anomaly score the $l1$ norm of the original image and the reconstructed image is used. 

Opposite to reconstruction-based anomaly detection methods as they are discussed above, in \citep{adgan:2020} adGAN, an alternative framework based on GANs, is proposed. The authors introduce two key components: fake pool generation and concentration loss. adGAN follows the structure of WGAN and  consists of a generator and discriminator. The WGAN is first trained with gradient penalty using healthy images only and after a number of iterations a pool of fake images is collected from the current generator. Then a discriminator  is retrained using the initial set of healthy data as well as the generated images in the fake pool with a concentration loss function. Concentration loss is a combination of the traditional WGAN loss function with a concentration term which aims to decrease the within-class distance of normal data. The output of the discriminator is considered as anomaly score. The method is applied to skin lesion detection and brain lesion detection. 
Two other methods that utilise discriminator outputs as anomaly score, however not tested for medical imaging, are ALOOC~\citep{alooc:2018} and fenceGAN \citep{fencegan:2019}.
In ALOOC \citep{alooc:2018}, the discriminator's probabilistic output is utilised as abnormality score. In their work an encoder-decoder is used for reconstruction while the discriminator tries to differentiate the reconstructed images from the original ones. An extension of the ALOOC algorithm, is the Old is Gold (OGN) algorithm which is presented in~\citep{uncertainty111:2020}. After training a framework similar to ALOOC, the authors finetune the network using two different types of fake images which are bad quality images and pseudo anomaly images. In this way they try to boost the ability of the discriminator to differentiate normal images from abnormal ones.

In \citep{fencegan:2019} the authors propose an encirclement loss that places the generated images at the boundary of the distribution and then use the discriminator in order to distinguish anomalous images. They propose this loss with the idea that a conventional GAN objective encourages the distribution of generated images to overlap with real images.

In ~\citep{medicalreview30:2020} an approach based on the ALOOC algorithm is proposed for the detection of fetal congenital heart disease. However, during training both normal and abnormal samples are available, which is one of the key differences compared to our approach where only healthy subjects are utilised. Furthermore, additional to the encoder-decoder and discriminator networks which are used in ALOOC, they use two additional noise models of the same architecture where the input is an image plus Gaussian noise ($\tilde{x}$) in order to make their encoder-decoder networks more robust to distortions.
In~\citep{uncertainty112:2019} a one-class generative adversarial network (OCGAN) is proposed for anomaly detection. OCGAN consists of two discriminators, a visual and a latent discriminator, a reconstruction network (denoising autoencoder) and a classifier. The latent discriminator learns to discriminate encoded real images and generated images randomly sampled from $\mathcal{U}\sim (-1,1)$, while the visual discriminator distinguishes real from fake images. Their classifier is trained using binary cross entropy loss and learns to recognise real images from fake images. Finally, in~\citep{uncertainty113:2018} a probabilistic framework is proposed which is based on a model similar to $\alpha$-GAN. The latent space is forced to be similar to standard normal distribution through an extra discriminator network, called latent discriminator similar to~\citep{alphagan:2017}. A parameterized data manifold is defined (using adversarial autoencoder) which captures the underlying structure of the inlier distribution (normal data) and a test sample is considered as  abnormal if its probability with respect to the inlier distribution is below a threshold. The probability is factorised with respect to local coordinates of the manifold tangent space.\\
A summary of the key features for the works above is given in Table~\ref{table:nonlin}.\\
To establish consistency between different related works we define $x$ as a test image, $\hat{x}$ as a reconstructed image, $D$ as a discriminator network, ($D_{H}$ as (intermediate) feature representation of a Discriminator network), $E$ as an encoder network (image space $\rightarrow${ latent space}), $D\text{e}$ as a decoder network (latent space back to image space), $G$ as a generator network (where input is a noise vector), $z$ as latent space representation and $\lambda$ as a fixed learning rate.

\begin{table}[ht!]
\caption{One-class anomaly detection using Generative Adversarial Networks}
\resizebox{\columnwidth}{!}{%
\begin{threeparttable}
\begin{tabular}{c c  c c}
\hline\hline
Reference & Approach& Anomaly score& Dataset \\ [0.5ex]
\hline
AnoGAN~\citep{anogan:2017} &reconstruction \& discrimination score &$(1-\lambda)\|x-G(z)\|+\lambda \|D_{H}(x)-D_{H}(G(z))\|$&OCT  \\
f-AnoGAN~\citep{fanogan:2019}& reconstruction \& discrimination score& $\|x-G(E(x)))\|^{2}+\lambda \|D_{H}(x)-D_{H}(G(E(x)))\|^{2}$&OCT   \\
Sparse-GAN~\citep{isbi:2020} &reconstruction error& $\|E(x)-E(De(E(x)))\|_{2}$& OCT  \\
AnoVAEGAN~\citep{baur2018deep:2018}&reconstruction error& $\|x-De(E(x))\|_1$&Brain \\
adGAN~\citep{adgan:2020}&discriminator score&$D(x)$&Digit/skin/Brain \\
*ALOOC~\citep{alooc:2018}&discriminator score&$D(De(E(x)))$&Generic Images/Video \\
*fenceGAN~\citep{fencegan:2019}&discriminator score  &$D(x)$& Generic Images\\ 
*OGN~\citep{uncertainty111:2020}&discriminator score&$D(De(E(x)))$&Generic Images/Video\\
*OCGAN~\citep{uncertainty112:2019}&discriminator/reconstruction score&$D(De(E(x)))$/$\|x-De(E(x))|^{2}$&Generic Images\\
*GPND~\citep{uncertainty113:2018}&probabilistic score&$p_{x}(x)$&Generic Images\\[1ex]
\hline
\end{tabular}
\begin{tablenotes}
\small
\item * Application field of these works as they are described in the original papers is not the Medical Imaging. 
\end{tablenotes}
\end{threeparttable}
}
\label{table:nonlin}
\end{table}

\section{Methods}
In order to detect anomalies in fetal ultrasound data, we build an end-to-end model which takes as input the whole image and produces an anomaly score together with an attention map in a unsupervised way. 

To achieve this, we build a GAN-based model, where the aim of the discriminator networks is to learn the salient features of the fetal images (\emph{i.e.}, heart area) during training. We use an auto-encoding generative adversarial network ($\alpha$-GAN) which makes use of discriminator information in order to predict the anomaly score. $\alpha$-GAN \citep{alphagan:2017}\citep{3dmrialphagan:2019} is a fusion of generative adversarial learning (GAN) and a variational autoencoder. It can be considered as autoencoder GAN combining the reconstruction power of an autoencoder with the sampling power of generative adversarial networks. It aims to overcome GAN instabilities during training, which leads to mode collapse while at the same time exploits the advantages of variational autoencoders, producing less blurry images. In $\alpha$-GAN two discriminators focus on the data and latent space respectively. An overview of the proposed architecture is given in Figure~\ref{fig:alphagan}

\begin{figure}[ht!]
\includegraphics[height=0.7\textwidth]{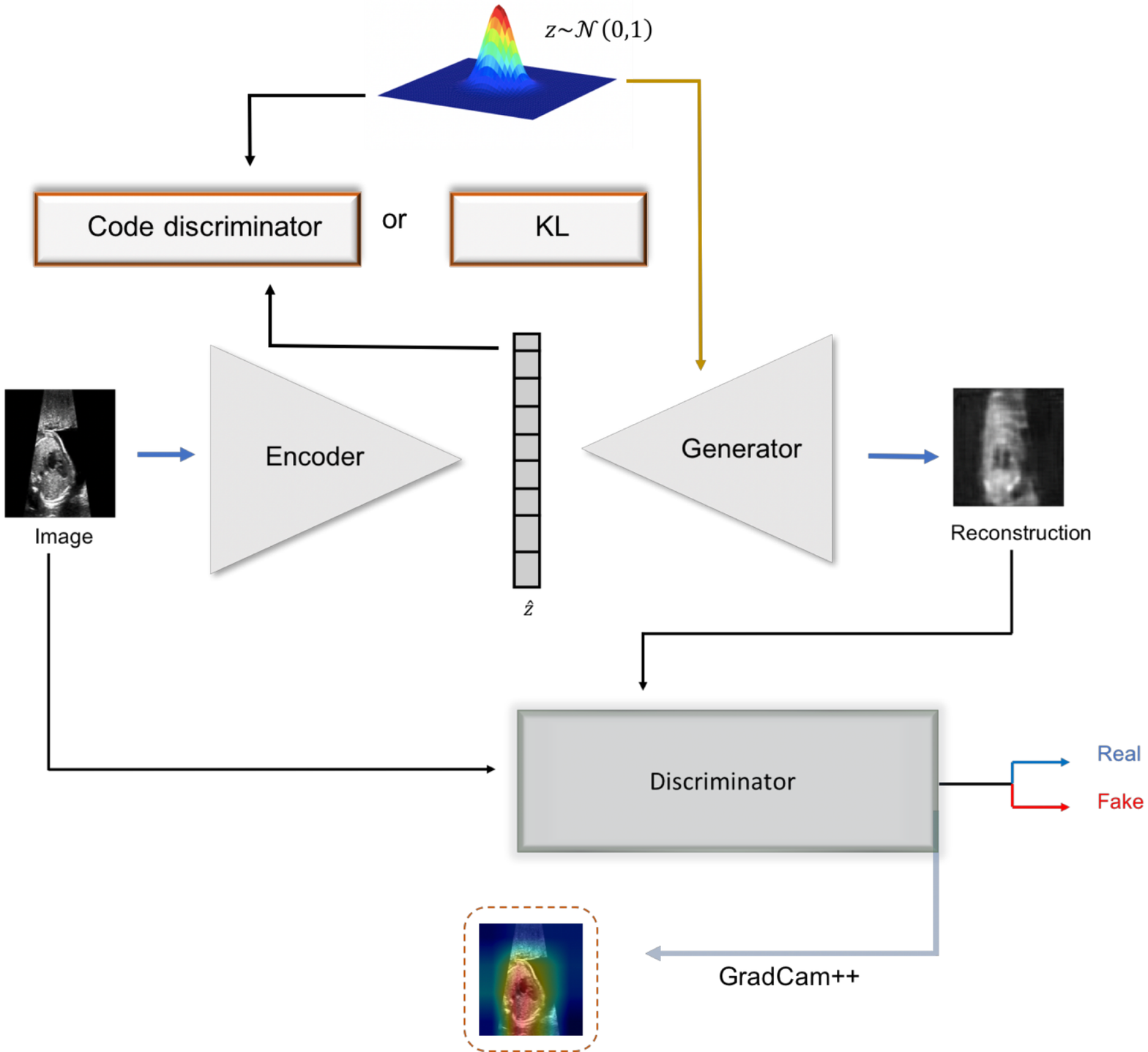}
\caption{Our proposed GAN-based model.}\label{fig:alphagan}
\end{figure}

\begin{figure}[ht!]
\alglanguage{pseudocode}
\renewcommand\figurename{ Algorithm}
\setcounter{figure}{\value{algorithm}}
\vspace{-0.5cm}%
\small
\hspace*{\algorithmicindent} \textbf{Input:} fetal ultrasound image $x$, parameter $\lambda$,  Number of Epochs: $N$ \\
\hspace*{\algorithmicindent} \textbf{Output:} Models:$E,G,D,LD$
\begin{algorithmic}[1]
\For {epoch 1 to $N$}
\State Update \textcolor{blue}{$E, G$} using Eqs.
\ref{eq:encoderloss}, \ref{eq:generatorloss}, \ref{eq:encgen}. $D, LD$ are fixed.
\Statex \textcolor{Fuchsia}{* $L_{\{.\}}$ indicates the loss function of each network}
\State $L_{E}\leftarrow{\lambda  \|x-\hat{x}\|_{1}+LD(\hat{z},1)}$
\State $L_{G}\leftarrow{\lambda  \|x-\hat{x}\|_{1}+D({\hat{x}},1)+D(\tilde{x},1)}$ 
\State $L_{E,G}\leftarrow{L_{E}+L_{G}}$
\State Update \textcolor{blue}{$D$} using Eq. \ref{eq:discrloss}. $E,G,LD$ are fixed.
\State  $L_{D}\leftarrow{D(x,1)+D(\hat{x},0)+D(\tilde{x},0)}$
\State Update \textcolor{blue}{$LD$} using Eq. \ref{eq:cdloss}. $E,G,D$ are fixed.
\State  $L_{LD}\leftarrow{LD(\hat{z},0)+LD(z,1)}$
\EndFor 
\Statex \textcolor{Fuchsia}{\small{* $G, D, LD, E$ indicates the corresponding outputs of each network. $1/0$ corresponds to  real/fake values.}}
\end{algorithmic} 
\caption{Training procedure of the proposed method.}
\label{alg:algGAN}
\end{figure}

We assume a generating process of real fetal cardiac images $x$ as $x \sim p^{*}_{x}$ and a random prior distribution $p_{z}$. Reconstruction, $\hat{x}$ , of an input image $x$ is defined as  $G(\hat{z})$ where $\hat{z}$ is a sample from the variational distribution $ \Large{q}_{E}$, \emph{i.e.}, $\hat{z} \sim \Large q_{E(z|x)}$. Furthermore, we define $z$ as a sample from a normal prior distribution $p_{z}$, \emph{i.e.}, $z \sim \mathcal{N}(0,I)$.

The encoder ($E$) is mapping each real sample $x$ from sample space $X$ to a point in the latent space $Z$, \emph{i.e.}, $E: X \rightarrow Z$. It consists of four blocks. Each block contains a Convolutional-Batch Normalisation layer followed by Leaky Rectified Linear Unit (LeakyReLU) activation, down-sampling the resolution of data by two in each block. Spectral Normalisation \citep{spectralnormalisation:2018}, \citep{attention2:2019}  a  weight normalisation method, is used after each convolutional layer. In the last block, after the convolutional layer, an attention gate is introduced  \citep{attention:2020}, \citep{attention2:2019}. The final layer of the encoder is a tangent layer. The dimension of the latent space is equal to $128$.

The generator synthesises images from latent space $Z$ back to the sample space $X$, \emph{i.e.}, $G: Z \rightarrow X$. The generator regenerates the initial image using four consecutive blocks of transposed convolution-batch normalisation-Rectified Linear Unit (ReLU) activation layers. The last layer is a Hyperbolic tangent (tanh) activation. Similar to encoder spectral normalisation, attention gate layers are  used.  

The discriminator ($D$) takes as input an image and tries to discriminate between real and fake images. The output of the discriminator is a probability for the input being a real or fake image. It consists of four blocks. Each block consists of Convolutional-Batch Normalisation-RELU layers. The last layer is a sigmoid layer. The discriminator treats $x$ as real images while the reconstruction from the encoder and samples from $p_{z}$, are considered as fake.

A latent discriminator is introduced in order to discriminate latent representations which are generated by the encoder network from samples of a standard Gaussian distribution. The latent code discriminator ($LD$) consists of four linear layers followed by a Leaky RELU activation. 
We randomly initialise the encoder, generator and latent code discriminator. The weights for the discriminator are initialised with a normal distribution $\mathcal{N} \sim (0,0.02)$. 
We train the architecture by first updating the encoder parameters by minimizing:
\begin{equation}
\begin{aligned}
\mathcal{L}_{\mbox{\small{E}}}= \EX_{p^{*}_{x}}[\lambda \times
\|x-\hat{x}\|_{1}+(-log(LD(\hat{z})))]
\label{eq:encoderloss}
\end{aligned}
\end{equation}
We define the generator loss as:
\begin{equation}
\begin{aligned}
\mathcal{L}_{\mbox{\small{G}}}= \EX_{p^{*}_{x}}[\lambda \times \|x-\hat{x}\|_{1}+(-log(D(G(\hat{z}))))]+ \EX_{p_{z}}[-log(D(G(z)))] 
\label{eq:generatorloss}
\end{aligned}
\end{equation}
Since we consider encoder and generator as one network the loss for the encoder-generator is:
\begin{equation}
\mathcal{L}_{E,G}=\mathcal{L}_{E}+\mathcal{L}_{G}
\label{eq:encgen}
\end{equation} where $\mathcal{L}_{E}$ and $\mathcal{L}_{G}$ are defined in Eqs. \ref{eq:encoderloss} and \ref{eq:generatorloss} respectively. The generator is updating twice compared to the encoder in order to stabilize the training procedure.\\\\
Then we minimise discriminator loss 
\begin{equation}
\mathcal{L}_{\mbox{\small{D}}}=\EX_{p^{*}_{x}}{[ -2*\log{\mbox{D}(x)}}-\log{(1-\mbox{D}(G(\hat{z})))]}+\EX_{p_{z}}{[-\log{(1-\mbox{D}(G(z)))}]}.
\label{eq:discrloss}
\end{equation}
Finally, we update the weights of latent code discriminator using
\begin{equation}
\mathcal{L}_{\mbox{\small{LD}}}=\EX_{p^{*}_{x}}{[-\log{(1-\mbox{LD}(\hat{z})})]}+\EX_{p_{z}}{[-\log({\mbox{LD}(z))}]}.  
\label{eq:cdloss}
\end{equation}

For the learning rate $\lambda$, we use value of $25$ after grid search. 

The training process of the $\alpha$-GAN model is described in algorithm \ref{alg:algGAN}. The networks are trained using the Adam optimizer. Encoder and Generator use the same learning rate, $\lambda$. The same learning rate is also utilised for discriminator and latent code discriminator.


We additionally replace the latent discriminator with an approximation of KL divergence. For a latent vector $\hat{z}$ of $M$ dimension we define KL divergence as~\citep{uncertainty114:2018}:
\begin{equation*}
KL(q(\hat{z}|x)||\mathcal{N}(0, I))\approx -\frac{M}{2}+\frac{1}{M}\sum_{i=1}^{M}\frac{s_{i}^{2}+m_{i}^2}{2}-log(s_{i}),  
\end{equation*}
where $m_{i}$ and $s_{i}$ is the mean and standard deviation of each component of the $M_{th}$ dimensional latent space. Performance in this configuration is subpar, thus we limit the discussion to results with the latent code discriminator.

Furthermore, we apply an analytic estimation of KL divergence using a one-class variational autoencoder (VAE-GAN) similar to~\citep{baur2018deep:2018}~\citep{uncertainty116:2016}. The VAE-GAN is trained using reconstruction error plus the KL divergence between the latent space ($\hat{z}$) and the normal distribution $p_{z}$.
For training the VAE-GAN, we first update the encoder and decoder networks as following:
\begin{equation*}
\mathcal{L}_{E}=\EX_{p^{*}_{x}}[\beta*\|x-\hat{x}\|_{p}]+KL(q(\hat{z}|x)||p_{z})
\end{equation*}
\begin{equation*}
\begin{aligned}
\mathcal{L}_{\mbox{\small{G}}}= \EX_{p^{*}_{x}}[\gamma \times \|x-\hat{x}\|_{p}+(-log(D(G(\hat{z}))))]+ \EX_{p_{z}}[-log(D(G(z)))] 
\end{aligned}
\end{equation*}
Finally, the discriminator is trained based on the: 
\begin{equation*}
\mathcal{L}_{\mbox{\small{D}}}=\EX_{p^{*}_{x}}{[ -2*\log{\mbox{D}(x)}}-\log{(1-\mbox{D}(G(\hat{z})))]}+\EX_{p_{z}}{[-\log{(1-\mbox{D}(G(z)))}]}.
\end{equation*}
where $\beta$, $\gamma$ are set to $10$ and $5$ respectively after grid search.

A ResNet18~\citep{uncertainty115:2016}-based architecture encoder and decoder/generator are utilised (with random initialisation). In the ResNet18 encoder/decoder architecture each layer consists of $4$ residual blocks and each block is $2-$ layer deep. We use the same discriminator as in $\alpha$-GAN.

The dimensions of the latent space are $128$. $p=2$ since we use the $l_{2}$ norm (\emph{i.e.}, mean square error).

All networks are implemented in Python using Pytorch, on a workstation with a NVIDIA Titan~X GPU.

\subsection{Anomaly detection score}
\label{sect:ascore}
In order to predict an anomaly score $s$, three different strategies are utilised. For an unseen image $x_{unseen}$ and its reconstructed image $\hat{x}_{unseen}$, we  utilise as baseline the reconstruction error which is defined as the $l2$ norm, \emph{i.e.}, $s_{rec}=\|x_{unseen}-\hat{x}_{unseen}\|_{2}^{2}$ between image and reconstructed image (residual). 

The second candidate for $s$ is the output of the discriminator. $D$ should give high scores for reconstructions of original, normal images, but low scores for abnormal images,  $s_{discr}=1-D(x_{unseen})$. 
Finally, we compute an anomaly score using a gradient-based method, GradCam++, ~\citep{gradcam++:2018}. 
Inspired by~\citep{adversarialdiscriminative:2020} ~\citep{attentionguidedanomaly:2020}~\citep{visuallyexplainedautoencoders:2020} we apply GradCam++ to the score of the discriminator with regards to the last rectified convolutional layer of the discriminator. This produces attention maps and is also valuable for the localisation of the pathology. The intuition of using attention maps for computing anomaly scores, is based on the hypothesis that after training the discriminator not only learns to discriminate between normal and abnormal samples but also learns to focus on relevant features in the image. Thus, specifically for HLHS, where the left artery is missing or is occluded compared to normal samples, a discriminator should identify and locate this difference. 
The GradCam++ is computed following:

Let $y$  be the logits of the last layer as they are derived from the discriminator network $D(x_{unseen})$. For the same operators  $(i, j)$ and $(a,b)$ applied to the feature map $A^{k}$ we compute weights:
\begin{equation}
w_{k}=\alpha_{ij}^{k} \mbox{RELU}(\frac{\partial y }{\partial A_{ij}^{k}}),
\label{eq:gradcampp_1}    
\end{equation} 
where the gradient weights $a_{ij}^{k}$ can be computed as:
\begin{equation}
\alpha_{ij}^k=\frac{\frac{\partial ^{2} y}{(\partial A_{ij}^{k})^{2}}}{2\frac{\partial ^{2} y}{(\partial A_{ij}^{k})^{2}}+\sum_{a}\sum_{b}A_{ab}^{k}\{{\frac{\partial ^{3} y}{(\partial A_{ij}^{k})^{3}}\}}}, 
\label{eq:gradcampp_2}
\end{equation}
and the saliency map (SM) is computed as a linear combination of the forward activation maps followed by a ReLU layer: 
\begin{equation}
SM_{ij}=RELU(\sum_{k}w_{k}A_{ij}^{k}).
\label{eq:gradcampp_4}    
\end{equation}
We then computed  the sum of the attention maps of image $x_{unseen}$ and its reconstruction from the Generator network, $\hat{x}_{unseen}$:
\begin{equation}
M=\mbox{SM}(D_{x_{unseen}})+\mbox{SM}(D_{\hat{x}_{unseen}})
\label{eq:gcampp}
\end{equation}
and finally computed the anomaly score $s_{attn}$ as
\begin{equation}
s_{attn} = \frac{\|M \times (x_{unseen}-\hat{x}_{unseen})\|_{2}^{2}}{\|M\|_{2}^{2}}
\label{eq:gcam3}
\end{equation}
To compute the anomaly score we encapsulate  the information of reconstruction~\citep{adversarialdiscriminative:2020}. Reconstruction of a normal image should be crisper compared to  reconstructions from an anomalous observation. Finally, we attempt to combine anomaly scores, such as $s_{rec}$ with $s_{discr}$. However, the anomaly detection performance does not improve noteworthily.

\subsection{Data}
The available dataset contains 2D ultrasound images of four-chamber cardiac views. These are standard diagnostic views according to~\citep{screening2015}. The images contain labelled examples from normal fetal hearts and hearts with Hypoplastic Left Heart Syndrome (HLHS)~\citep{hlh1:2019} from the same clinic, using exclusively an Aplio i800 GI system for both groups to avoid systematic domain differences. HLHS is a birth defect that affects normal blood flow through the heart. It 
affects a number of structures on the left side of the heart that do not fully develop. 

Our dataset consists of $2317$ 4-chamber view images for which $2224$ cases are normal and $93$ are abnormal cases. Healthy control view planes have been automatically extracted from examination screen capture videos using a Sononet network \citep{sononet:2017} and manual cleaning from visually trivial classification errors. A set of HLHS view planes that would resemble a 4-chamber view in healthy subjects has been extracted with the same automated Sononet pipeline. Another set has been manually extracted from the examination videos by a fetal cardiologist and 38 cases that are not within 19+0 - 20+6 weeks or show a mix of pathologies have been rejected.  

\begin{figure}[ht!]
\centering
\includegraphics[width=.95\linewidth]{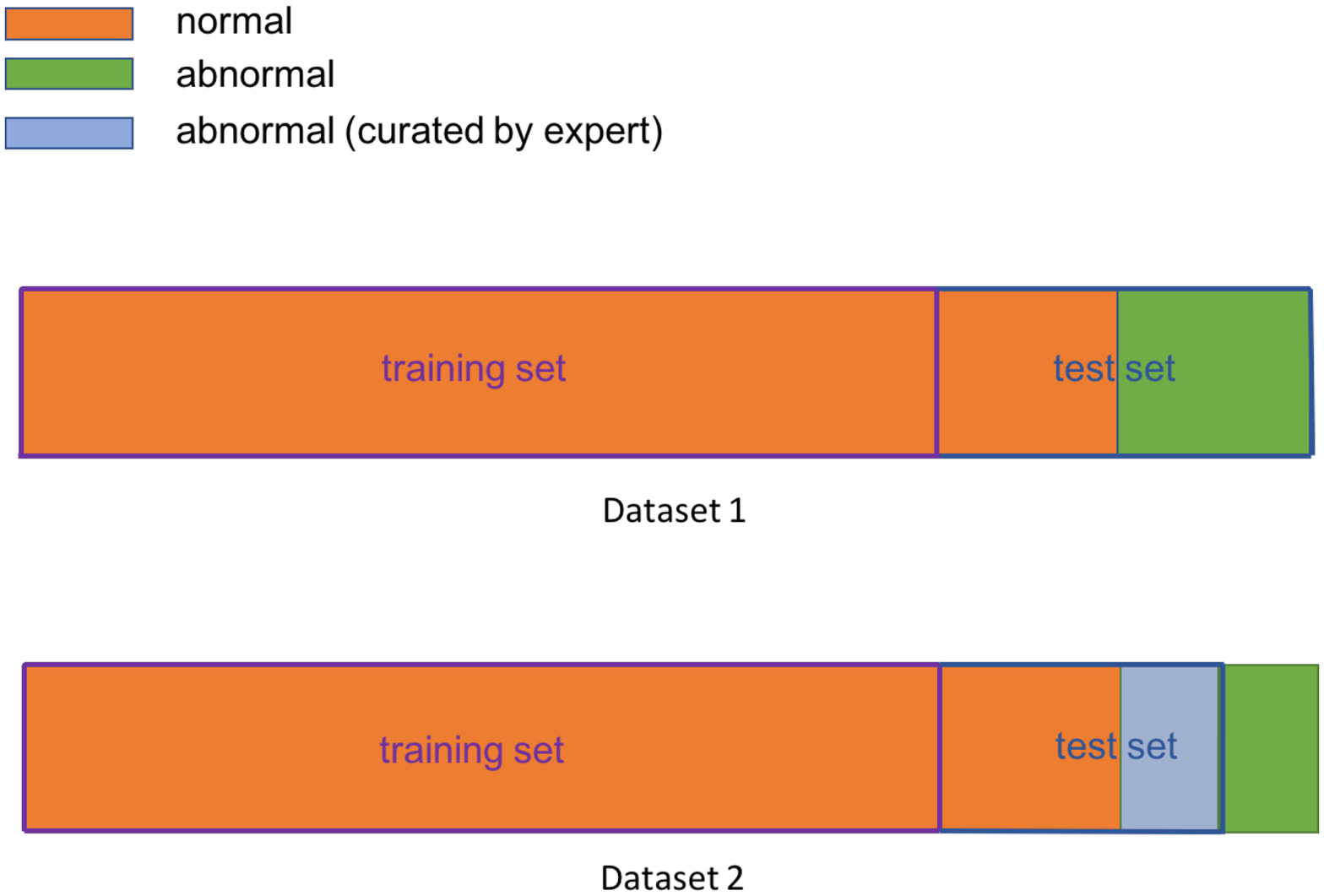}
\caption{Graphical description of Dataset $1$ and Dataset $2$}
\label{fig:descriptdataset}
\end{figure}

For training, $2131$ 4-chamber view images, which are considered as normal cases are used. During training, only images from normal fetuses are used. For testing, two different datasets are derived for three different testing scenarios:

For $\mathbf{dataset_{1}}$(Figure ~\ref{fig:descriptdataset}) we use 4-chamber views from all available HLHS cases, extracted by Sononet and cleaned from gross classification errors; in total $93$ cases. Further $93$ normal cases have been randomly selected from the remaining test split of the healthy controls and added to this dataset. HLHS cases are challenging for Sononet, which has been trained only on healthy views. Thus, in HLHS cases, it will only select views that are close to the feature distribution of healthy 4-chamber views, which are not necessarily the views a clinician would have chosen. 
For $\mathbf{dataset_{2}}$(Figure ~\ref{fig:descriptdataset}), we use the $93$ normal cases from $dataset_{1}$ and the expert-curated HLHS images from the remaining, nonexcluded $53$ cases. For each of these cases $1$ to $4$ different view planes have been identified as clinically conclusive. 
With this dataset we perform two different subject-level experiments: a) selecting one of the four frames randomly and b) using all of the 177 clinically selected views in these 53 subjects and fusing the individual abnormality scores to gain a subject-level assessment. We also evaluate per-frame anomaly results. 
The images are rescaled to $64 \times 64$ and normalised to a $[0,1]$ value range. No image augmentation is used.

\section{Evaluation and Results}
We evaluate our algorithm both quantitatively as well as qualitatively. The capability of the proposed method to localise the pathology is also examined.

\subsection{Quantitative analysis}
For evaluation purposes, the anomaly score is computed as described in Section~\ref{sect:ascore}. For $\alpha$-GAN and VAE-GAN we use $s_{attn}$, $s_{rec}$ and $s_{discr}$ as anomaly scores as they are presented in Section~\ref{sect:ascore}.

For comparison with the state-of-the-art we train four algorithms:
convolutional autoencoder (CAE) \citep{cae1:2015, cae2:2011},  One-class Deep Support Vector Data Description (DeepSVDD)~\citep{deepsvdd:2018} and f-AnoGAN~\citep{fanogan:2019}.

Deep Convolutional autoencoder (DCAE) ~\citep{cae1:2015, cae2:2011} is also trained as a baseline. For training, MSE loss is utilised. For DCAE and One-class DeepSVDD we use the same architectures as the ones used for the CIFAR10 dataset in the original work~\citep{deepsvdd:2018}.
Reconstruction error, \emph{i.e.}, $\|x-De(E(x))\|_{2}$, is defined as anomaly score ($s_{DCAE}$).

Deep Support Vector Data Description (DeepSVDD)~\citep{deepsvdd:2018} 
computes the hypersphere of minimum volume that contains every point in the training set.  By minimising the sphere's volume, the chance of including points that do not belong to the target class distribution is minimised.
Since in our case all the training data belongs to one class (negative class-healthy data) we focus on~\citep{deepsvdd:2018} .
Let $f$ be the network function of the deep neural network with $L$ layers and  $\theta^{l}$ the weights's parameters of the $l_{th}$ layer. We denote  the center of the hypersphere as $o$. The objective of the network is to minimize the loss which is defined as:
\begin{equation*}
\mathcal{L}_{SVDD} = \min_{\theta}\frac{1}{N}\sum_{i=1}^{N}  \|f(x)-o\|^{2}+\frac{\lambda}{2}\sum_{l=1}^{L}\|\theta^{l}\|^{2}.    
\end{equation*}
The center $o$ is set to be the mean of outputs which is obtained at the initial forward pass.
The anomaly score ($s_{svdd}$) is then defined at inference stage as the distance between a new test sample to the center of the hyper-sphere, \emph{i.e.}, $\|f(x)-o\|^{2}$

f-AnoGAN~\citep{fanogan:2019} is described in Section~\ref{sotamethods}. We were not able to  successfully train f-AnoGAN using the same networks as we used for $\alpha$-GAN, hence we utilise similar networks and an identical training framework as  described in~\citep{fanogan:2019}. We follow the \textit{$izi_{f}$} training procedure for the encoder network. As anomaly detection score ($s_{anogan}$) a combination of $L2$ residual loss between the image and its reconstruction and the $L2$ norm of the discriminator's features of an intermediate layer is utilised as it is defined in Table~\ref{table:nonlin}.

\begin{figure}[ht!]
 \centering
\subfloat[][\label{fig:roc}]{\includegraphics[height=6cm]{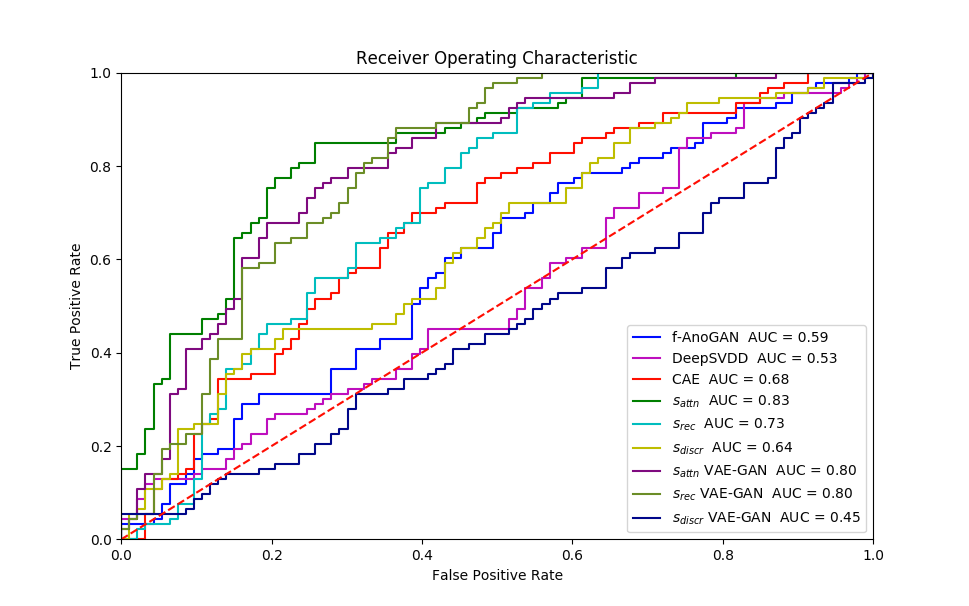}}
\subfloat[][\label{fig:distrib}]{\includegraphics[height=6cm]{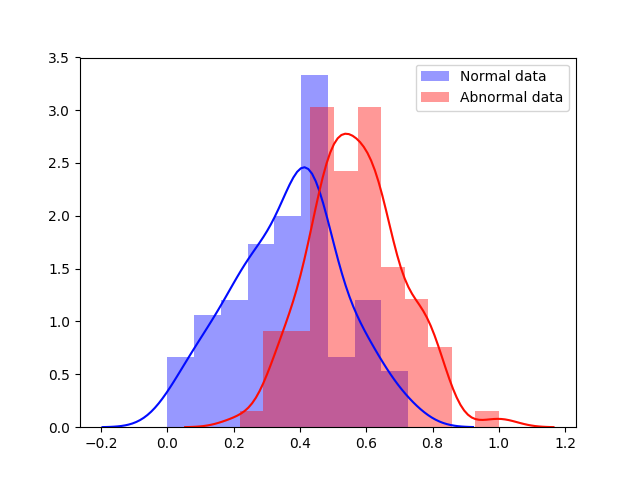}}\\
\subfloat[][\label{fig:conf}]{\includegraphics[height=6cm]{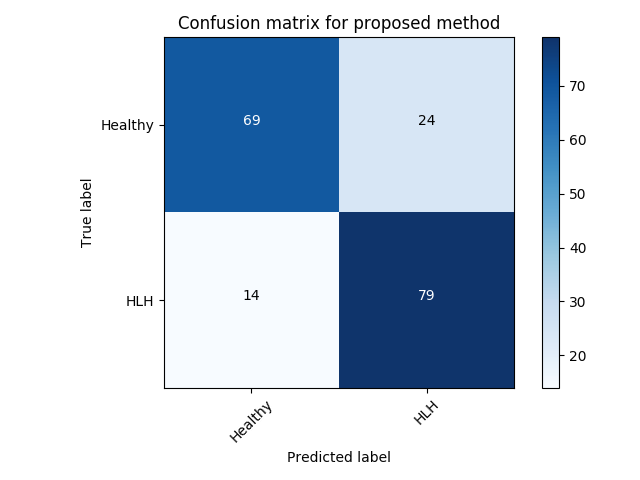}}
\subfloat[][\label{fig:confvae}]{\includegraphics[height=6cm]{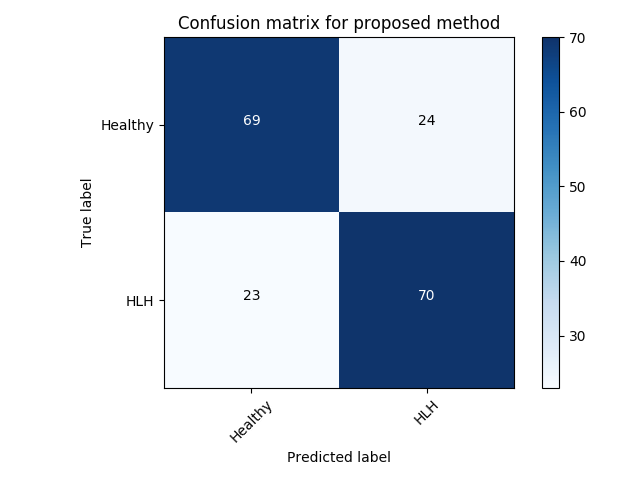}}
\caption{(a) ROC-AUC curves in Exp. 1; (b) Distribution of normal/abnormal score values for the $\alpha$-GAN model with  $s_{attn}$ as  anomaly score (c) Confusion matrix for the best performing run of the proposed $\alpha$-GAN (d) Confusion matrix for the best performing run of the VAE-GAN.  
This figure focuses on the results of the best performing initialisation from  five experiments with $\alpha$-GAN (or VAE-GAN) while Table~\ref{tab:auc_score_new} shows average metrics.}
\label{fig:results}
\end{figure}

In all algorithms the latent dimension is chosen as $128$. We run all experiments $5$ times using different random seeds~\citep{equalgan:2018}. We report the average precision, recall at the Youden index of the receiver operating characteristic (ROC) curves as well as the average corresponding area under curve (AUC) of the $5$ runs of each experiment. 
Furthermore, we apply the DeLong's test~\citep{DeLong:1988} to obtain z-scores and p-values in order to test how statistically different the AUC curve of the proposed model compared to the corresponding curves of the state-of-the-art models (CAE and DeepSVDD, f-AnoGAN and VAE-GAN) is.
We perform four different experiments:

\textbf{Experiment 1} uses $dataset_{1}$ and aims to evaluate general, frame-level outlier detection performance, including erroneous classifications and fetuses below the expected age range. In Table~\ref{tab:auc_score_new}, the best performing model based on AUC score is the $\alpha$-GAN method using $s_{attn}$ as anomaly score which achieves an average of $0.82\pm 0.012$ AUC. The $\alpha$-GAN model achieves the best precision score. However, regarding F1 score and Recall VAE-GAN outperforms $\alpha$-GAN with $0.88$ and $0.78$ respectively.
DeepSVDD shows the best specificity at $0.76$. Figure~\ref{fig:results} shows the ROC for the best performing (AUC, F1) initialisation and the distribution of normal 
and abnormal scores for the best model of $\alpha$-GAN at the Youden index. We present confusion matrices for the $\alpha$-GAN and the VAE-GAN models in Figure~\ref{fig:conf} and Figure~\ref{fig:confvae}. For normal cases both models achieve similar classification performance. However, for identifying abnormal cases $\alpha$-GAN seems to have an advantage.

Based on the DeLong's test, for Exp. 1, for the average scores (of five experiments), $\alpha$-GAN compared to f-AnoGAN yields $z =-5.22$ and $p =1.80e-07$. Similarly, the values for $\alpha$-GAN compared to CAE are $z =-4.82$ and $p=1.37e-06$. Finally, comparing $\alpha$-GAN and DeepSVDD results in $z = -6.49$ and $p = 8.52e-11$. Since $p<0.01$ for all comparisons, we can assume that $\alpha$-GAN performs significantly better than the state-of-the-art when applied to fetal cardiac ultrasound screening for HLHS. 
Comparing $\alpha$-GAN  with VAE-GAN the values are $z=-1.21$ and $p=0.22$ which does not indicate a significant difference between AUC curves.
As can be seen from the results, the GAN-based methods achieve better performance for detecting HLHS.

\begin{table}[ht!]
\centering
\resizebox{\columnwidth}{!}{%
\begin{tabular}{ |p{5.5cm}||p{2cm}|p{2cm}|p{2cm}|p{2cm}|p{2cm}|}
 \hline
 \multicolumn{6}{|c|}{Quantitative performance scores}\\
 \hline
Method& Precision & Recall& Specificity &F1 score & AUC\\
 \hline
CAE \small \citep{deepsvdd:2018} &$0.65 \pm 0.027$&$0.64 \pm 0.061$&$0.65 \pm 0.074$&$0.64 \pm 0.061$&$0.65 \pm 0.016$\\
DeepSVDD \small \citep{deepsvdd:2018} &$0.67 \pm 0.106$&$0.37 \pm 0.258$&$\mathbf{0.76} \pm 0.260$&$0.41 \pm 0.150$&$0.53 \pm 0.039$\\
f-AnoGAN \small \citep{fanogan:2019}&$0.58 \pm 0.022$&$0.58 \pm 0.130$&$0.59 \pm 0.097$&$0.57 \pm 0.072$&$0.57 \pm 0.039$\\
$s_{rec}$ (VAE-GAN)&$0.69 \pm 0.018$&$\mathbf{0.88} \pm 0.060$&$0.61 \pm 0.057$&$\mathbf{0.78} \pm 0.015$&$0.78 \pm 0.010$\\
$s_{discr}$ (VAE-GAN)&$0.75 \pm 0.220$&$0.29 \pm 0.360$&$0.75 \pm 0.360$&$0.27 \pm 0.230$&$0.42 \pm 0.027$\\
$s_{attn}$ (VAE-GAN)&$0.72 \pm 0.014$&$0.83 \pm 0.043$&$0.68 \pm 0.037$&$0.77 \pm 0.014$&$0.79 \pm 0.008$\\
$s_{rec}$ ($\alpha$-GAN) &$0.64 \pm 0.017$&$0.87 \pm 0.054$&$0.50 \pm 0.038$&$0.74 \pm 0.024$&$0.71 \pm 0.029$\\
$s_{discr}$ ($\alpha$-GAN) &$0.65 \pm 0.056$&$0.51 \pm 0.240$&$0.70 \pm 0.205$&$0.53 \pm 0.170$&$0.61 \pm 0.067$\\
$s_{attn}$ ($\alpha$-GAN)&$\mathbf{0.73} \pm 0.026$&$0.82 \pm 0.068$&$0.70 \pm 0.059$&$0.77 \pm 0.029$&$\mathbf{0.82} \pm 0.012$\\
\hline
\end{tabular}
}
\caption{\label{tab:auc_score_new} Anomaly detection performance for Exp. 1 using $dataset_{1}$. Best performance in bold.}
\end{table}

\begin{figure}[ht!]
 \centering
\subfloat[][\label{fig:roc1}]{\includegraphics[height=6cm]{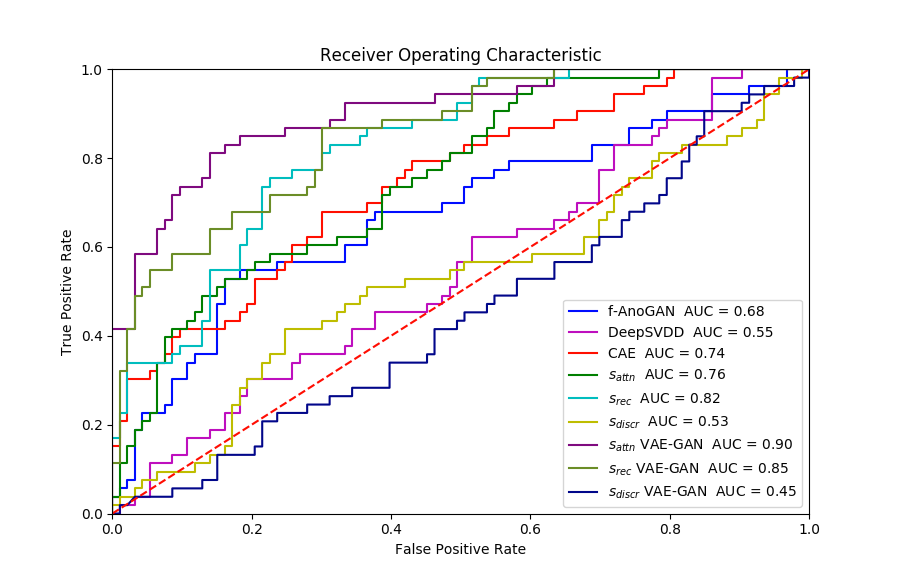}}
\subfloat[][\label{fig:distrib1}]{\includegraphics[height=6cm]{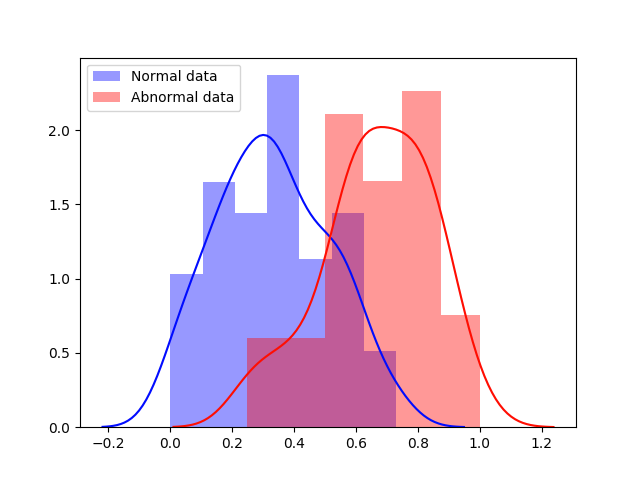}}\\
\subfloat[][\label{fig:conf1}]{\includegraphics[height=6cm]{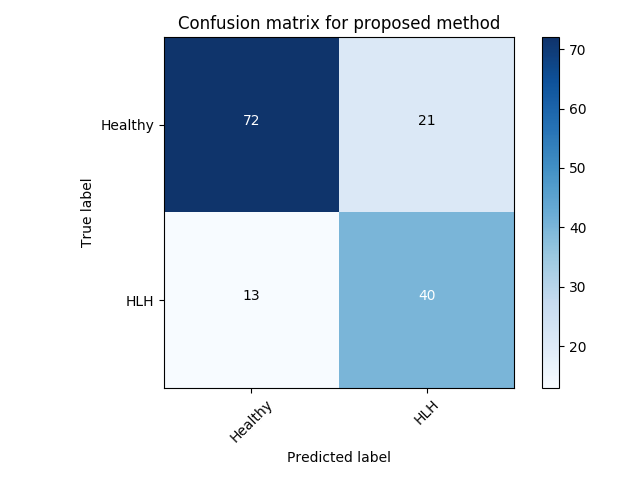}}
\subfloat[][\label{fig:conf11}]{\includegraphics[height=6cm]{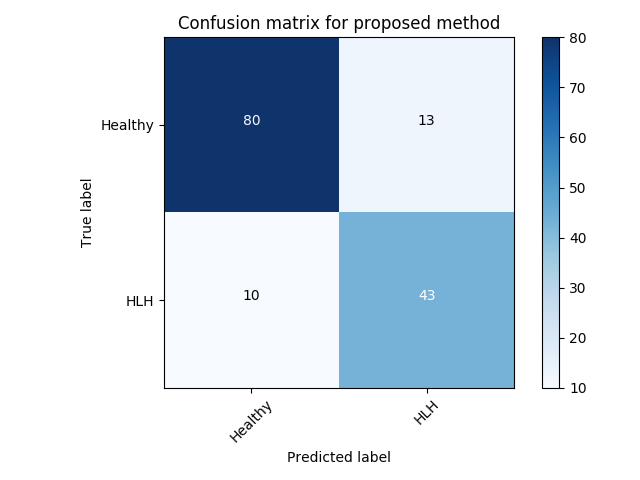}}
\caption{$dataset_{2}$, Exp. 2: (a) ROC-AUC curves in Exp. 2; (b) Distribution of normal/abnormal score values for the VAE-GAN model with  $s_{attn}$ as  anomaly score (c) Confusion matrix for the best performing run  using $s_{rec}$ of the proposed $\alpha$-GAN. (d)  Confusion matrix for the best performing run using $s_{attn}$ of the VAE-GAN. This figure focuses on the results of the best performing initialisation from  five experiments with $\alpha$-GAN (or VAE-GAN) while Table~\ref{tab:auc_score_new1} shows average metrics.}
\label{fig:results_dataset2}
\end{figure}

\textbf{Experiment 2} uses $dataset_{2}$ for specific disease detection capabilities with expert-curated, clinically conclusive 4-chamber views for 53 HLHS cases. We choose one of the relevant views per subject randomly.  Table~\ref{tab:auc_score_new1} summarises these results.  VAE-GAN has the highest AUC, F1, precision and specificity scores using $s_{attn}$ as anomaly score. Also, we note from Figure~\ref{fig:conf1} and Figure~\ref{fig:conf11} that the VAE-GAN method misclassified less HLHS cases while achieving better performance for confirming normal cases. Average F1 score is $0.89$. 
Figure~\ref{fig:results_dataset2} shows ROC, anomaly score distribution and confusion matrices  at the Youden index of this experiment. 

\begin{table}[ht!]
\centering
\resizebox{\columnwidth}{!}{%
\begin{tabular}{ |p{5.5cm}||p{2cm}|p{2cm}|p{2cm}|p{2cm}|p{2cm}|}
 \hline
 \multicolumn{6}{|c|}{Quantitative performance scores}\\
 \hline
Method& Precision & Recall& Specificity &F1 score & AUC\\
 \hline
CAE \small \citep{deepsvdd:2018} &$0.63 \pm 0.095$&$0.56 \pm 0.120$&$0.78 \pm 0.130$&$0.57 \pm 0.025$&$0.72 \pm 0.015$\\
DeepSVDD \small \citep{deepsvdd:2018} &$0.39 \pm 0.016$&$0.80 \pm 0.160$&$0.28 \pm 0.160$&$0.52 \pm 0.032$&$0.49 \pm 0.038$\\
f-AnoGAN \small \citep{fanogan:2019}&$0.56 \pm 0.077$&$0.52 \pm 0.097$&$0.75 \pm 0.140$&$0.53 \pm 0.041$&$0.63 \pm 0.043$\\
$s_{rec}$ (VAE-GAN)&$0.64 \pm 0.067$&$0.80 \pm 0.060$&$0.74 \pm 0.078$&$0.71 \pm 0.020$&$0.84 \pm 0.009$\\
$s_{discr}$ (VAE-GAN)&$0.36 \pm 0.220$&$0.56 \pm 0.450$&$0.46 \pm 0.430$&$0.34 \pm 0.205$&$0.39 \pm 0.037$\\
$s_{attn}$ (VAE-GAN)&$\mathbf{0.71} \pm 0.046$&$0.85 \pm 0.038$&$\mathbf{0.80} \pm 0.058$&$\mathbf{0.77} \pm 0.016$&$\mathbf{0.89} \pm 0.009$\\
$s_{rec}$($\alpha$-GAN) &$0.59 \pm 0.050$&$\mathbf{0.81} \pm 0.060$&$0.66 \pm 0.010$&$0.68 \pm 0.015$&$0.79 \pm 0.030$\\
$s_{discr}$($\alpha$-GAN) &$0.48 \pm 0.100$&$0.51 \pm 0.280$&$0.61 \pm 0.280$&$0.43 \pm 0.110$&$ 0.53 \pm 0.030$\\
$s_{attn}$($\alpha$-GAN) &$0.59 \pm 0.098$&$0.76 \pm 0.150$&$0.66 \pm 0.180$&$0.64 \pm 0.037$&$0.77 \pm 0.046$\\
\hline
\end{tabular}
}
\caption{\label{tab:auc_score_new1} Anomaly detection performance using $dataset_{2}$ for Exp. 2. Best performance in bold.}
\end{table}

\begin{figure}[ht!]
 \centering
\subfloat[][\label{fig:roc2}]{\includegraphics[height=6cm]{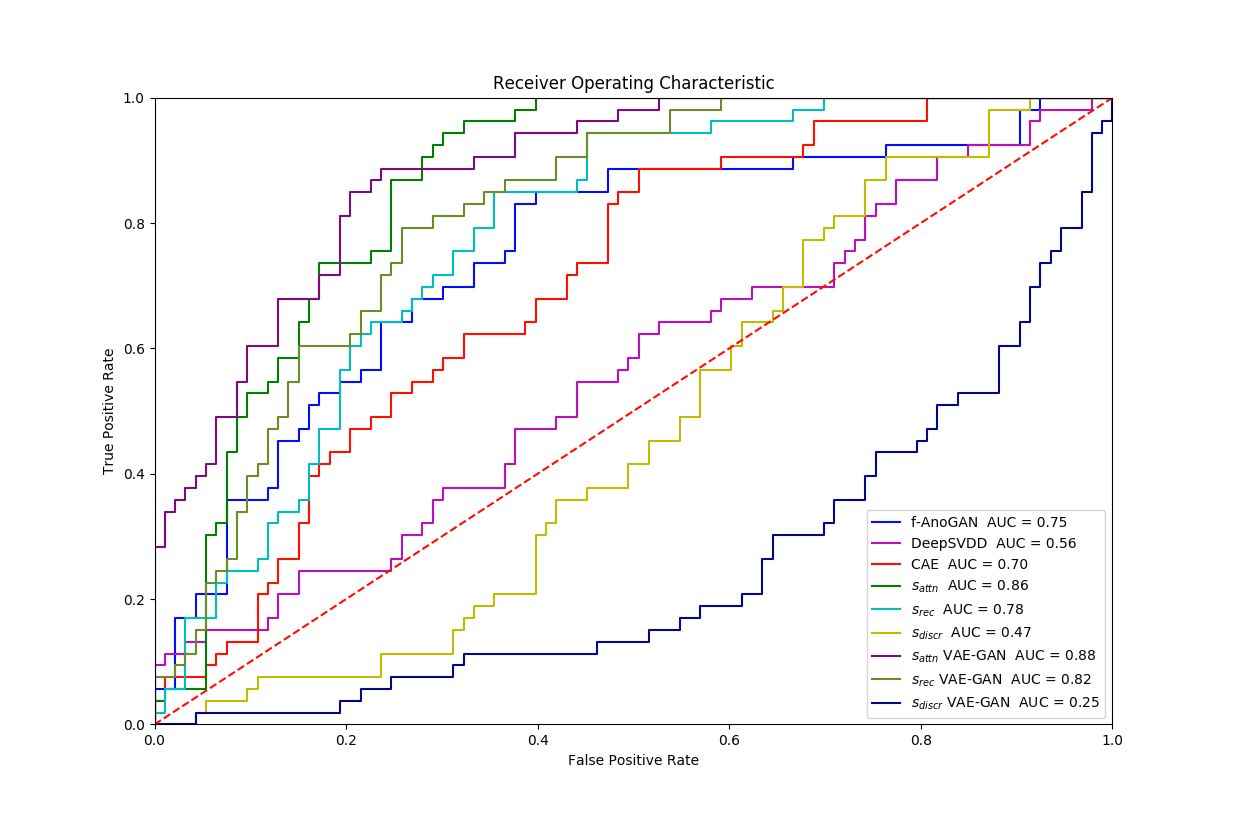}}
\subfloat[][\label{fig:distrib2}]{\includegraphics[height=6cm]{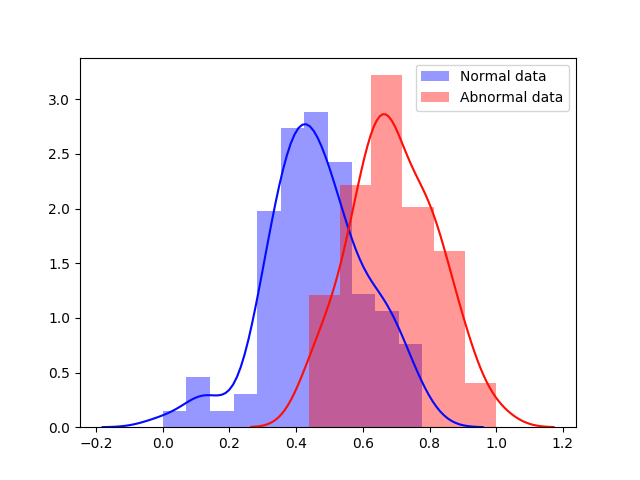}}\\
\subfloat[][\label{fig:conf2}]{\includegraphics[height=6cm]{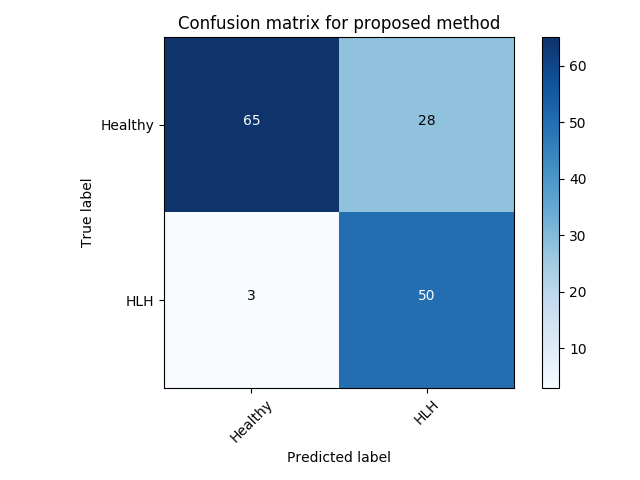}}
\subfloat[][\label{fig:conf22}]{\includegraphics[height=6cm]{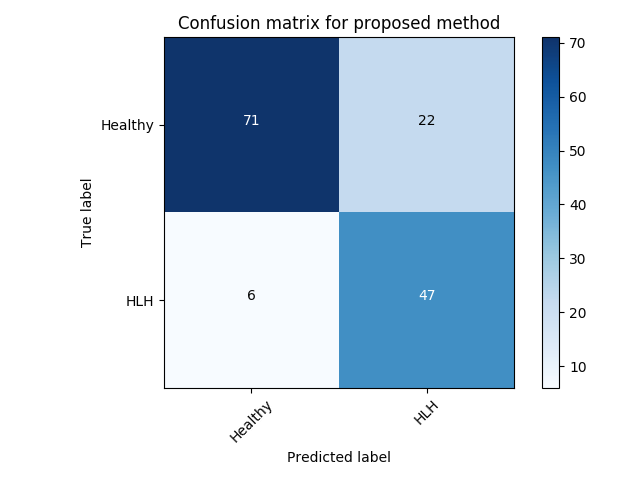}}
\caption{$dataset_{2}$, Exp. 3: (a) ROC-AUC curves in Exp. 3; (b) Distribution of normal/abnormal score values for the VAE-GAN model with  $s_{attn}$ as  anomaly score (c) Confusion matrix for the best performing run of the proposed $\alpha$-GAN (d) Confusion matrix for the best performing run of the VAE-GAN. This figure focuses on the results of the best performing initialisation from five experiments with $\alpha$-GAN (or VAE-GAN)) while Table~\ref{tab:auc_score_new2} shows average metrics.}
\label{fig:results_dataset3}
\end{figure}

\textbf{Experiment 3} uses $dataset_{2}$ and is similar to Exp. 2 except that we take all clinically identified views for each subject into account. We average the individual anomaly scores for each frame, depending on the number of frames that are available per subject. VAE-GAN achieves a better AUC score with $0.86$ compared to $0.84$ of $\alpha$-GAN as can be seen in Table \ref{tab:auc_score_new2}. However, as can be seen from the confusion matrices (best performing initialisation), $\alpha$-GAN shows a better true positive rate at the cost of a higher number of false positives (Figure \ref{fig:conf2}). This configuration might be preferred in a clinical setting since it reduces the number of missed cases at the cost of a slightly higher number of false referrals.

\begin{table}[ht!]
\centering
\resizebox{\columnwidth}{!}{%
\begin{tabular}{ |p{5.5cm}||p{2cm}|p{2cm}|p{2cm}|p{2cm}|p{2cm}|}
 \hline
 \multicolumn{6}{|c|}{Quantitative performance scores}\\
 \hline
Method& Precision & Recall& Specificity &F1 score & AUC\\
 \hline
CAE \small \citep{deepsvdd:2018} &$0.51 \pm 0.061$&$0.80 \pm 0.136$&$0.54 \pm 0.150$&$0.61 \pm 0.018$&$0.70 \pm 0.024$\\
DeepSVDD \small \citep{deepsvdd:2018} &$0.42 \pm 0.063$&$0.69 \pm 0.312$&$0.39 \pm 0.311$&$0.47 \pm 0.140$&$0.48 \pm 0.038$\\
f-AnoGAN \small \citep{fanogan:2019} &$0.55 \pm 0.029$&$0.79 \pm 0.067$&$0.62 \pm 0.068$&$0.64 \pm 0.016$&$0.74 \pm 0.013$\\
$s_{rec}$ (VAE-GAN)&$0.60 \pm 0.029$&$0.87 \pm 0.049$&$0.67 \pm 0.056$&$0.71 \pm 0.014$&$0.81 \pm 0.076$\\
$s_{discr}$ (VAE-GAN)&$0.37 \pm 0.150$&$0.98 \pm 0.400$&$0.032 \pm 0.39$&$0.53 \pm 0.021$&$0.14 \pm 0.034$\\
$s_{attn}$ (VAE-GAN)&$\mathbf{0.66} \pm 0.036$&$0.88 \pm 0.035$&$\mathbf{0.74} \pm 0.050$&$\mathbf{0.75} \pm 0.014$&$\mathbf{0.86} \pm 0.017$\\
$s_{rec}$ ($\alpha$-GAN) &$0.57 \pm 0.041$&$0.86 \pm 0.091$&$0.62 \pm 0.098$&$0.68 \pm 0.022$&$0.78 \pm 0.019$\\
$s_{discr}$ ($\alpha$-GAN) &$0.42 \pm 0.035$&$0.89 \pm 0.110$&$0.28 \pm 0.155$&$0.57 \pm 0.067$&$0.48 \pm 0.017$\\
$s_{attn}$ ($\alpha$-GAN) &$0.62 \pm 0.040$&$\mathbf{0.92} \pm 0.100$&$0.67 \pm 0.069$&$0.73 \pm 0.024$&$0.84 \pm 0.018$\\
\hline
\end{tabular}
}
\caption{\label{tab:auc_score_new2} Anomaly detection performance on subject level for $dataset_{2}$ and Exp. 3. Best performance in bold.}
\end{table}

\textbf{Experiment 4} is similar with the Exp. 3 except that we evaluate frame-level performance in Table~\ref{tab:auc_score_new3}. 
VAE-GAN is again better in terms of precision and AUC performance. However, similar to Exp. 3 $\alpha$-GAN has an advantage when recognising the cases with pathology at a cost of a higher false positive rate.

\begin{table}[ht!]
\centering
\resizebox{\columnwidth}{!}{%
\begin{tabular}{ |p{5.5cm}||p{2cm}|p{2cm}|p{2cm}|p{2cm}|p{2cm}|}
 \hline
 \multicolumn{6}{|c|}{Quantitative performance scores}\\
 \hline
Method& Precision & Recall& Specificity &F1 score & AUC\\
 \hline
CAE \small \citep{deepsvdd:2018} &$0.80 \pm 0.026$&$0.57 \pm 0.081$&$0.71 \pm 0.075$&$0.66 \pm 0.051$&$0.67 \pm 0.020$\\
DeepSVDD \small \citep{deepsvdd:2018} &$0.86 \pm 0.100$&$0.09 \pm 0.030$&$\mathbf{0.96} \pm 0.025$&$0.15 \pm 0.053$&$0.44\pm 0.025$\\
f-AnoGAN \small \citep{fanogan:2019} &$0.82 \pm 0.041$&$0.56 \pm 0.070$&$0.75 \pm 0.095$&$0.66 \pm 0.040$&$0.66 \pm 0.013$\\
$s_{rec}$ (VAE-GAN)&$0.82 \pm 0.023$&$0.74 \pm 0.062$&$0.69 \pm 0.073$&$0.77 \pm 0.024$&$0.77 \pm 0.009$\\
$s_{discr}$ (VAE-GAN)&$0.80 \pm 0.130$&$0.03 \pm 0.007$&$0.99 \pm 0.008$&$0.05 \pm 0.012$&$0.37 \pm 0.047$\\
$s_{attn}$ (VAE-GAN)&$\mathbf{0.86} \pm 0.016$&$0.78 \pm 0.051$&$0.76 \pm 0.046$&$0.82 \pm 0.023$&$\mathbf{0.82} \pm 0.023$\\
$s_{rec}$ ($\alpha$-GAN) &$0.80 \pm 0.016$&$0.80 \pm 0.032$&$0.62 \pm 0.051$&$0.80 \pm 0.012$&$0.75 \pm 0.017$\\
$s_{discr}$ ($\alpha$-GAN) &$0.71 \pm 0.060$&$0.72 \pm 0.300$&$0.38 \pm 0.320$&$0.66 \pm 0.180$&$0.48 \pm 0.055$\\
$s_{attn}$ ($\alpha$-GAN) &$0.82 \pm 0.030$&$\mathbf{0.85} \pm 0.110$&$0.64 \pm 0.094$&$\mathbf{0.83} \pm 0.047$&$0.81 \pm 0.018$\\
\hline
\end{tabular}
}
\caption{\label{tab:auc_score_new3} Anomaly detection performance using $dataset_{2}$ in Exp. 4 for evaluation per frame. Best performance in bold.}
\end{table}

\begin{figure}[ht!]
\centering
\subfloat[][\label{fig:roc3}]{\includegraphics[height=6cm]{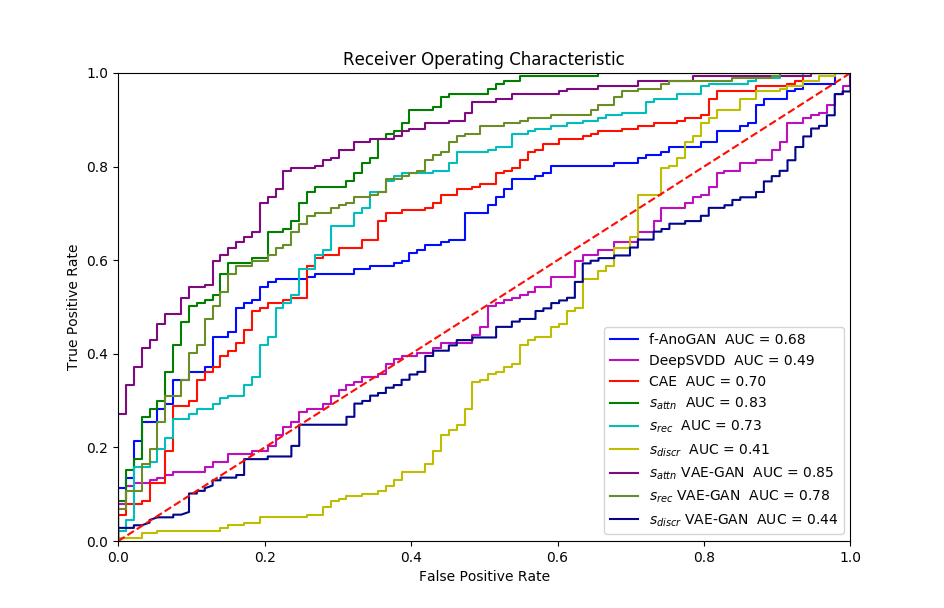}}
\subfloat[][\label{fig:distrib3}]{\includegraphics[height=6cm]{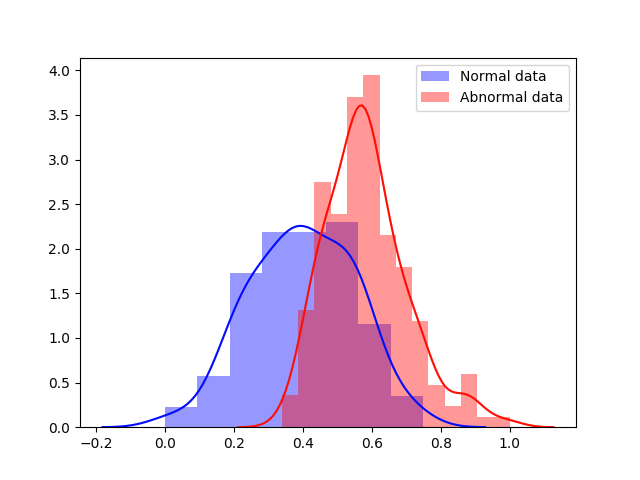}}\\
\subfloat[][\label{fig:conf3}]{\includegraphics[height=6cm]{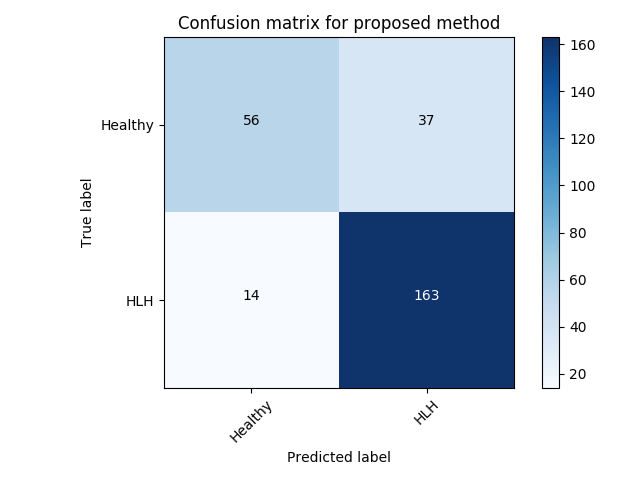}}
\subfloat[][\label{fig:conf33}]{\includegraphics[height=6cm]{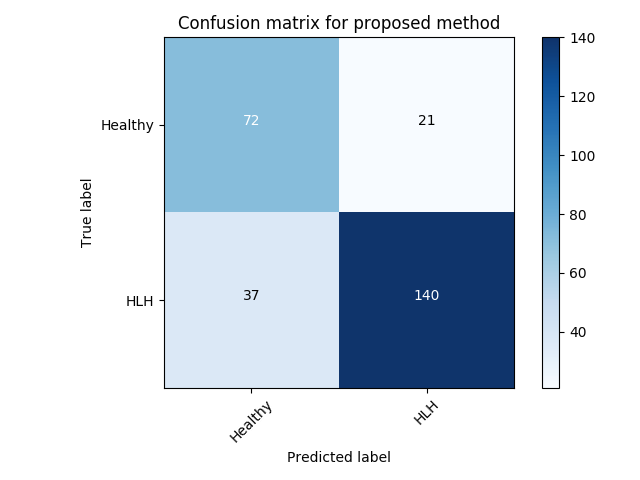}}
\caption{$dataset_{2}$, Exp. 4: (a) ROC-AUC curves in Exp. 4; (b) Distribution of normal/abnormal score values for the $\alpha$-GAN model with  $s_{attn}$ as  anomaly score (c) Confusion matrix for the best performing run of the proposed $\alpha$-GAN. (d) Confusion matrix for the best performing run of the VAE-GAN. This figure focuses on the results of the best performing initialisation from  five experiments with $\alpha$-GAN (or VAE-GAN) while Table~\ref{tab:auc_score_new3} shows average metrics.}
\label{fig:results_dataset4}
\end{figure}

\newpage
\subsection{Qualitative analysis}
\begin{figure}[ht!]
\centering
\includegraphics[width=0.7\textwidth]{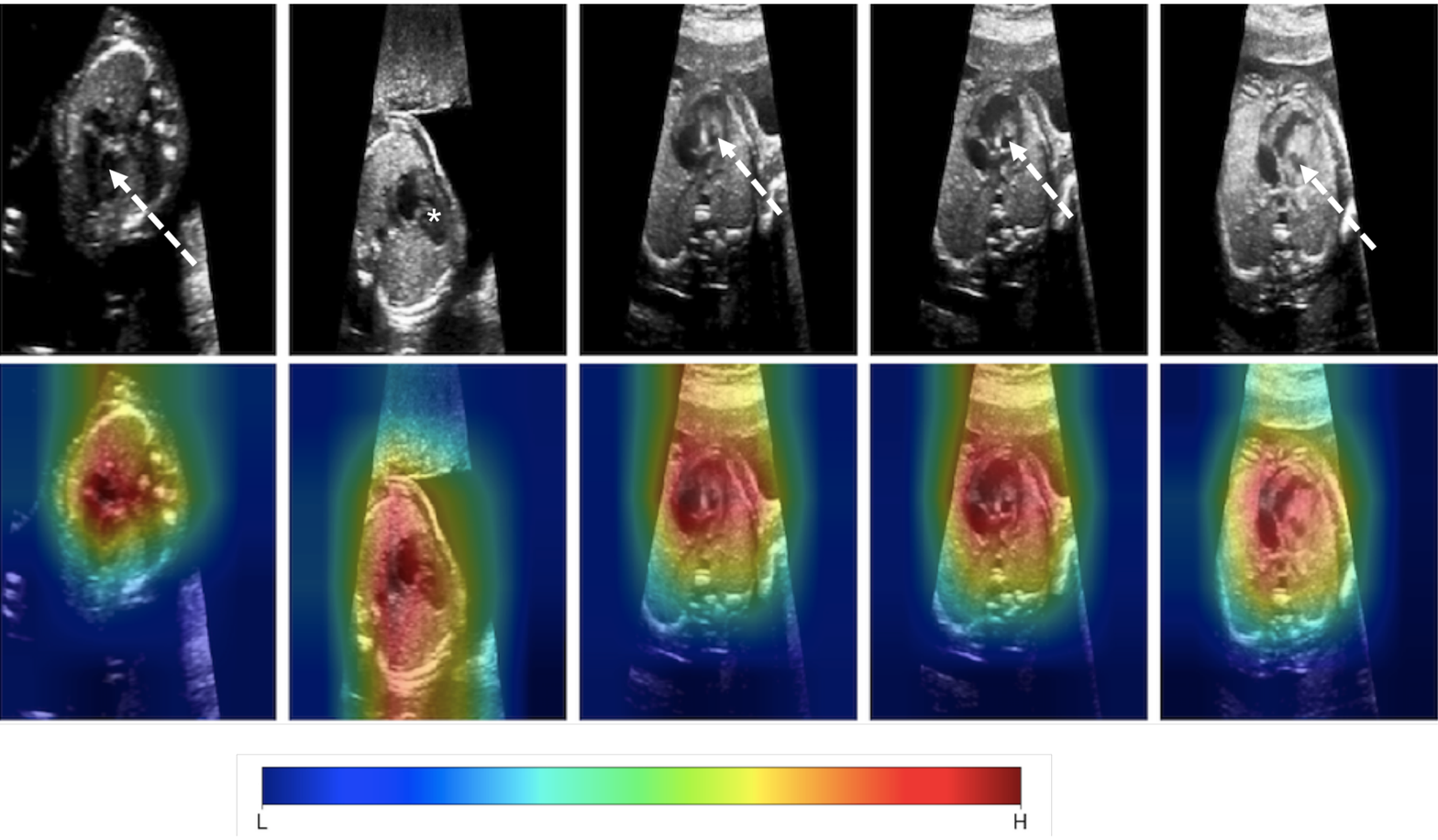}
\caption{Top row: Pathological subjects Bottom row:  GradCam++ visualisation of attention maps using $\alpha$-GAN (Exp. 1).\\ *= dominant RV with no visible LV cavity, solid white arrow = deceptively normal-looking LV, dashed white arrow = globular, hypoplastic LV}
\label{fig:camunhealthy}
\end{figure}

\begin{figure}[ht!]
 \centering
\subfloat[][]{\includegraphics[width=.65\linewidth]{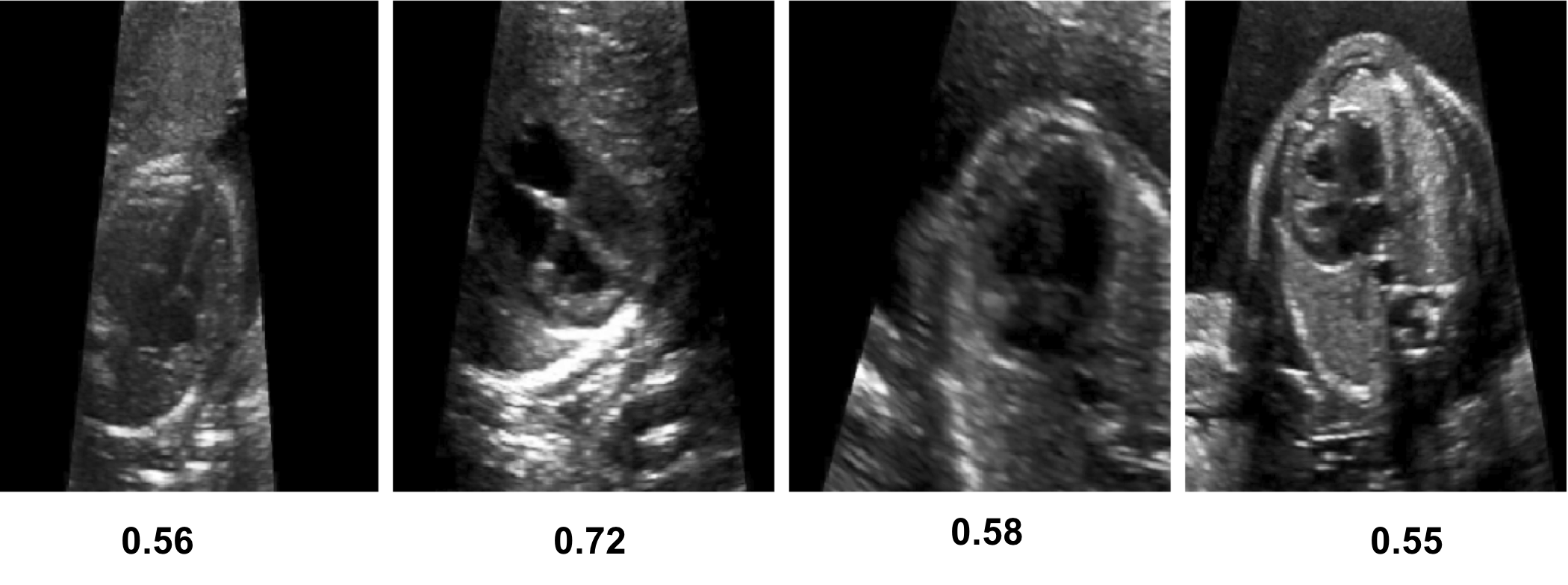}\label{fig:falsepositives}}\\
\subfloat[][]{\includegraphics[width=.65\linewidth]{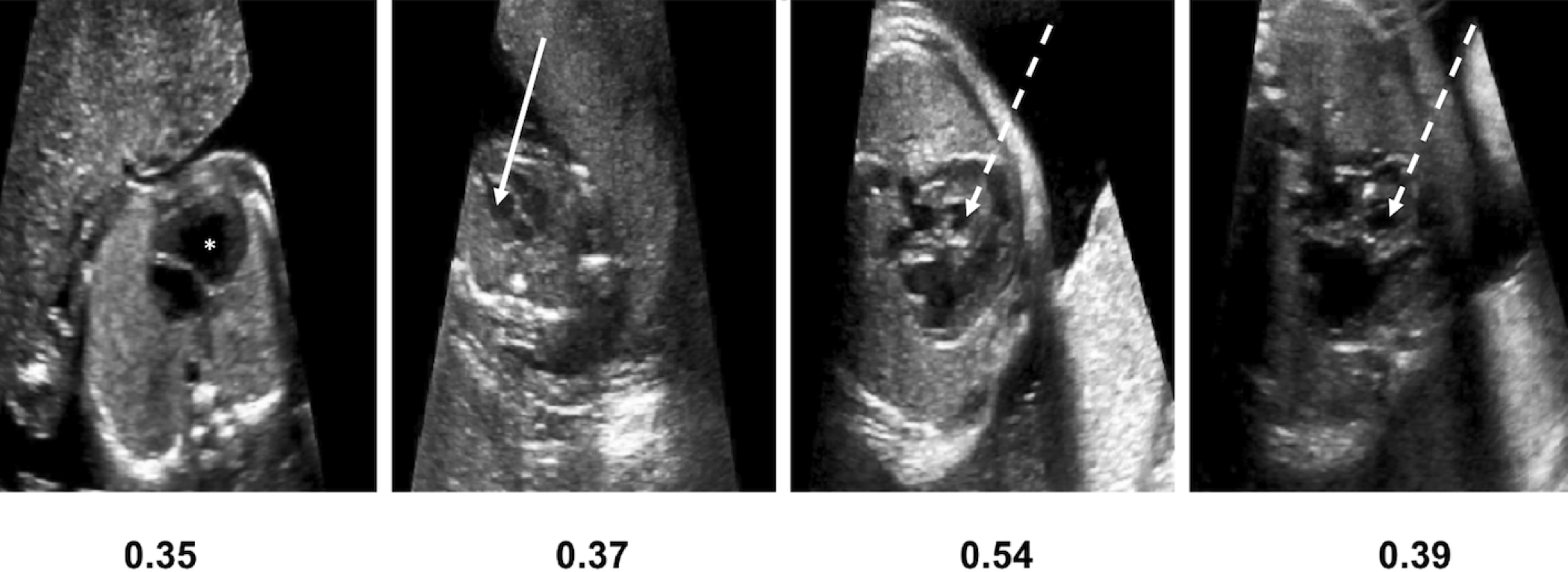}\label{fig:falsenegatives}}
\caption{(a) Examples of False Positive along with the anomaly scores $s_{attn}$ (b) False Negative cases along with the anomaly scores $s_{attn}$ (Exp. 1). *= dominant RV with no visible LV cavity, solid white arrow = deceptively normal-looking LV, dashed white arrow = globular, hypoplastic LV. Low Signal-to-Noise Ratio (SNR)}
\label{fig:plot_fpfn}
\end{figure}

In order to evaluate the ability of the algorithm to localise anomalies, we plot the class activation maps as they are derived from the proposed model. We present results from abnormal cases in $dataset_{1}$ (Exp.1) Figure~\ref{fig:camunhealthy}. In the abnormal cases, attention focus exactly in the area of heart. As a consequence, anomaly scores in such cases are higher compared to normal cases and correctly indicating the anomalous cases. All anomaly scores are normalised in the range of $[0,1]$. There are  cases that our algorithm fails to classify correctly. Either they are abnormal and they are classified as normal (False Negative-FN) or they are healthy and identified as anomalous (False Positive-FP). 
In Figure~\ref{fig:plot_fpfn} examples for False Positive cases are presented alongside False Negative cases. Bad image reconstruction quality is a limiting factor. For instance, in some reconstructions either a part of the heart is missing (left or right ventricle/ atrium) or the shape of the heart is quite different from a normal heart (\emph{e.g.}, a very ``long'' ventricle). As a consequence, not only the reconstruction error is high, but also the attention mechanism focuses in this area, since it is recognised (by the network) as anomalous. Consequently, the total anomaly score is high. In fewer examples the signal-to-noise ratio (SNR) is low, \emph{i.e.}, images are blurry, and so the network fails to reconstruct the images at all. Furthermore, in the False Positive examples Figure~\ref{fig:falsepositives}, from clinical perspective, the angle is not quite right, so it makes the ventricles look shorter than they are. This confuses the model, forcing the discriminator's attention to indicate this area as anomalous. Another point which is very interesting to highlight, is that there are cases where some frames are very difficult, even for experts. Such an example is given in Figure~\ref{fig:falsenegatives}, where although the second image from left belongs to an abnormal subject, the specific frame appears normal at the first glance. Such cases also highlight limitations of single-view approaches. In practice, all relevant frames showing the four chamber view could be processed with our method and a majority vote could regarding referral be calibrated on a ROC curve.  

All the above plots and comparisons utilise the top-$1$ performing experiment among all the runs of the experiments for $\alpha$-GAN.

\section{Discussion}
Our results are  promising and confirm that automated anomaly detection can work in fetal 2D ultrasound as shown on the example of HLHS. For this pathology we achieve an average accuracy of $0.81$ AUC, improving significantly the detection rate of front-line-of-care sonographers during screening, which is often below 60\%~\citep{chew2007population}. 
However, there are  open issues.

False negative rates are critical for clinical diagnosis and downstream treatment. In a clinical setting, a method with zero false negative predictions would be preferred, \emph{i.e.}, a method that \emph{never} misses an anomaly, but potentially predicts a few false positives. Assuming that the false positive rate of such an algorithm is significantly below the status quo, the benefits for antenatal detection and potentially better postnatal outcomes would outweigh the costs. 
Of course, an algorithm with a 100\% false positive rate is also not desirable, hence calibration on the ROC must be performed. 

A key aspect of the proposed algorithm is the ability of the discriminator to highlight decisive areas in images. In order to achieve this, it is necessary to produce good reconstructions of normal images. However, reconstruction quality can be limited, depending on the given sample. A larger dataset could provide a mitigation strategy for this. Furthermore, alternative ways for visualising attention could be explored for disease-specific applications such as implicit mechanisms of attention like attention gates~\citep{attention:2020}.

Although we have experimented with different type of noise (e.g Uniform) and various augmentation techniques (e.g horizontal flip, intensity changes) we did not notice an improvement in anomaly detection performance. However, a further investigation of other augmentation techniques should be done.


Moreover, it would be interesting to explore the sensitivity of our method for other sub-types of congenital heart disease. Intuitively, accuracy of a general anomaly detection method should be similarly high for other syndromes that affect the morphological appearance of the fetal four-chamber view. HLHS has a particularly grossly abnormal appearance. There are a lot of other CHD examples with a subtly abnormal four chamber view that would probably be much harder to detect even for human experts. Additionally, in practice, confounding factors may bias anomaly detection methods towards more obvious outliers, while subtle signs of disease or indicators encoded in other dimensions like the spatio-temporal domain may still be missed.

Finally, robust time-series analysis is still a challenging fundamental research question and we are looking forward to extending our method to full video sequences in future work.


\section{Conclusion}
In this paper we attempt to consider the detection of congenital heart disease as a one-class anomaly detection problem, learning only from normal samples. The proposed unsupervised architecture shows promising results and achieves better performance compared to existing state-of-the-art image anomaly detection methods. However, since clinical practice requires highly reliable anomaly detection methods,  more work will need to be done to avoid false positives to mitigate patient stress and strain on healthcare systems and false negatives to prevent missed diagnoses.

\section*{Acknowledgements} EC was supported by an EPSRC DTP award. TD was supported by an NIHR Doctoral Fellowship. We thank the volunteers and sonographers from routine fetal screening at St. Thomas' Hospital London. This work was supported by the Wellcome Trust IEH Award [102431] for the Intelligent Fetal Imaging and Diagnosis project (\url{www.ifindproject.com})  and EPSRC EP/S013687/1. The study has been granted NHS R\&D and ethics approval, NRES ref no = 14/LO/1086.
The research was funded/supported by the National Institute for Health Research (NIHR) Biomedical Research Center based at Guy's and St Thomas' NHS Foundation Trust, King's College London and the NIHR Clinical Research Facility (CRF) at Guy's and St Thomas'. Data access only in line with the informed consent of the participants, subject to approval by the project ethics board and under a formal Data Sharing Agreement. The views expressed are those of the author(s) and not necessarily those of the NHS, the NIHR or the Department of Health.

\bibliography{sample}

\end{document}